\documentclass[twocolumn]{autart}

\usepackage{graphicx}
\usepackage{braket}
\usepackage{amsmath,amssymb,amsfonts,latexsym,mathrsfs}
\usepackage{comment}
\usepackage{xcolor}
\usepackage{cite}

\newcommand{\Tr}{\operatorname{Tr}}

\newcommand{\A}{\mathbf{A}}

\renewcommand{\H}{\mathbf{H}}

\begin{document}

\begin{frontmatter}
\title{Robust Quantum Control in Closed and Open Systems: Theory and Practice\thanksref{footnoteinfo}}

\thanks[footnoteinfo]{This paper was not presented at any IFAC 
meeting. Corresponding author E.~A.~Jonckheere.}

\author[CAW]{C. A. Weidner}\ead{c.weidner@bristol.ac.uk},
\author[EAR]{E. A. Reed} \ead{ree93038@ttu.edu},
\author[JM]{J. Monroe} \ead{jonathan.t.monroe@boeing.com},
\author[BS]{B. Sheller}\ead{shellerba.math@gmail.com},
\author[SO]{S. O'Neil}\ead{sonei@usc.edu},
\author[EM]{E. Maas} \ead{eliavmaas@gmail.com},
\author[EAJ]{E. A. Jonckheere}\ead{jonckhee@usc.edu},
\author[FCL]{F. C. Langbein}\ead{frank@langbein.org},
\author[SGS]{S. G. Schirmer}\ead{s.m.shermer@gmail.com}
\address[CAW]{Quantum Engineering Technology Laboratories, H. H. Wills Physics Laboratory and Department of Electrical and Electronic Engineering, University of Bristol, Bristol BS8 1FD, United Kingdom}
\address[EAR]{Department of Electrical and Computer Engineering, Texas Tech University, Lubbock, TX 79409, USA}
\address[JM]{Boeing Research \& Technology, Huntington Beach, CA 92647, USA}
\address[BS]{Department of Computer Science, Drake University, 2507 University Ave, Des Moines, IA 50311 USA}
\address[SO]{Department of Electrical and Computer Engineering, University of Southern California, Los Angeles, California 90089, USA}
\address[EM]{Department of Electrical and Computer Engineering, University of Southern California, Los Angeles, California 90089, USA}
\address[EAJ]{Department of Electrical and Computer Engineering, University of Southern California, Los Angeles, California 90089, USA}
\address[FCL]{School of Computer Science and Informatics, Cardiff University, Cardiff CF24 4AG, United Kingdom}
\address[SGS]{Faculty of Science and Engineering (Physics), Swansea University, Swansea, SA2 8PP, United Kingdom}
          
\begin{keyword}
quantum control; robust control; quantum information; quantum systems
\end{keyword}

\begin{abstract}
Robust control of quantum systems is an increasingly relevant field of study amidst the second quantum revolution, but there remains a gap between taming quantum physics and robust control in its modern analytical form that culminated in fundamental performance bounds. 
With certain exceptions such as quantum optical systems that can be modelled as linear stochastic differential equations, quantum systems are not amenable to linear, time-invariant, measurement-based robust control techniques, and thus novel gap-bridging techniques must be developed. This survey is written for control theorists to provide a review of the current state of quantum control and outline the challenges faced in trying to apply modern robust control to quantum systems.  We present issues that arise when applying classical robust control theory to quantum systems, typical methods used by quantum physicists to explore such systems and their robustness, as well as a discussion of open problems to be addressed in the field.  We focus on general, practical applications and recent work to enable control researchers to contribute to advancing this burgeoning field.
\end{abstract}
\end{frontmatter}

\section{Introduction}\label{sec:intro}

As quantum technologies continue to mature, their development must transition from proofs-of-principle to well-engineered systems with numerous commercial applications in computing, sensing, and networking. This transformation of quantum technologies into the real-world application space requires the development of \emph{robust} means to control and manipulate quantum systems. Quantum control theory has been developed to the point where several textbooks~\cite{cong,dalessandro,dirr} and comprehensive review papers~\cite{Ball2020,Rabitz2010,doherty2000quantum,Petersen2010,Glaser2015,Koch_2016} have been written on the subject. While classical robust control has been extensively studied and is well-understood~\cite{Limebeer,Postlethwaite,Zhou}, rigorous development of robust control protocols for quantum mechanical systems remains a challenging field of research. This is because classical control methods targeting fundamental limitations, worst-case $H^\infty$ performance, multi-variable gain and phase margins, structured singular value, etc. cannot be readily applied to quantum systems in general.

Although some areas, such as quantum optics, lend themselves easily to a classical, if non-commutative, stochastic control formulation~\cite{Petersen2010,Petersen2008,quantum_feedback,Petersen2013}, there are three main stumbling blocks in the adaptation of automatic control techniques to general quantum problems. The first one is \emph{controlling at the edge of stability}.  Indeed, coherent quantum systems are purely oscillatory, hence marginally stable. Under decoherence, quantum systems can be stabilized, but most such systems will ultimately lose their quantum properties with time, and steady-state solutions to these systems are often devoid of any quantum advantage. The challenge of robust quantum control is often (but not universally) a compromise between stability and quantum advantage. Furthermore, progress in decoherence-based state preparation~\cite{Carvalho2001,Verstraete2009} and bath engineering~\cite{Harrington2022,Kapit2017} has not heavily leveraged robust control theory. The second major issue is that of bilinear versus linear systems. Fundamentally, quantum control deals with bilinear systems~\cite{Elliott}. To make classical robust control adaptable to quantum control, one approach is to make the bilinear system linear by time-invariant bias fields or piecewise time-invariant fields and recover linear control schemes. However, the work in \cite{bilinear_constant_input} shows that closed-loop pole placement in bilinear systems subject to constant gain control differs from classical linear feedback control systems. Nevertheless, one great advantage of this constant gain control scheme is that it allows the study of the \emph{feedback} properties~\cite{Safonov_Laub_Hartmann} of quantum systems without explicitly feeding back measurements, hence circumventing the back-action of such measurements. It can be shown that one can even obtain a quantum advantage from, e.g., coherent feedback control~\cite{Mabuchi2012}. The third discrepancy is that classical control usually quantifies performance in terms of frequency-response inequalities while quantum state transfer, gate optimization, etc. require that the specifications are met at a \emph{precise}, preferably short, terminal time $t_f$.

Therefore, more research is needed into the theoretical underpinnings of robust quantum control \emph{sidestepping measurements}, as well as practical application and eventual implementation of quantum controls into real systems. The following overarching questions remain to be answered: Can a quantum system ever be inherently robust, especially in the absence of stability? What are the fundamental device limitations established by quantum robust control protocols? Will researchers ever be able to move past the current noisy, intermediate-scale quantum (NISQ) era and build useful, scalable, and robust devices that are promised by the second quantum revolution? While this remains to be seen, some hope can be offered by the success of related applications that rely on quantum phenomena and control such as nuclear magnetic resonance (NMR) and magnetic resonance imaging (MRI) (see, e.g.~\cite{grape,NMR_2,NMR_1,Chuang_2005,wimperis1994broadband}, among many others). If researchers can see a coherent signal from the many protons contained in the water that makes up (most of) the human body a warm, wet, complex chemical environment, there may yet be hope for large-scale quantum computers.

As a result of the relative immaturity of robust quantum control, the barrier to entry into the field is quite high, as there are few good references for researchers in related fields to gain an overview of the state-of-the-art and open questions in the area. This survey attempts to fill this gap.  Some aspects of (robust) quantum control are not considered in this survey, as they have been covered elsewhere~\cite{Petersen2013}.  Specifically, this survey will not cover linear quantum systems, a special class of quantum optical systems that can be mapped onto linear, time-invariant (LTI) systems. An introduction to robust control for these systems can be found in~\cite{Petersen2010,Petersen2013,robust_closed_loop_quantum}, which provide excellent reviews, and current research in this domain can be found in, e.g.,~\cite{Wang_2023}.  

Measurement-based control, coherent feedback control, and Lyapunov control are also beyond the scope of this survey.  These are covered by existing survey and tutorial papers and textbooks~\cite{quantum_feedback,quantum_modeling_control,wiseman_milburn}.  In a similar vein, this survey will not cover the so-called \emph{measurement problem} and other philosophical aspects of quantum mechanics~\cite{RMP_measurement}.  While these aspects can sometimes be leveraged for control design, they do not play a significant role in the control paradigms considered in this survey. Finally, we will not consider \emph{adaptive} control, which endeavors to identify the potentially uncertain parameters, and techniques such as the \emph{spectator} approach~\cite{nicest_spectator_paper,spectator_qubits}. Beyond these restrictions, our approach is to be as general as possible through a discussion of closed and open quantum systems, robust control challenges framed in the context of classical control, and current methods for finding optimal controls and notions of robustness used in practice.

This survey is organized as follows: Sec.~\ref{sec:survey} presents a brief survey of broadly related literature, Sec.~\ref{sec:basic} discusses key issues that arise when applying classical robust control methods to quantum systems, particularly those stemming from the differences between classical linear systems and bilinear systems describing quantum mechanics. Current techniques for finding robust quantum controllers are described in Sec.~\ref{sec:struct}, and current avenues of research in robust quantum control in Sec.~\ref{sec:current}, effectively surveying recent developments in the field with a focus on enabling the reader to understand some of the ideas and methods currently used in the nascent field of robust quantum control as it is developed.

\section{General Overview of the Literature}\label{sec:survey}

This section does not aim to provide a thorough overview of the research in the field of quantum control. Rather, we highlight some previous work in the areas of quantum optics, quantum landscape control, homodyne detection, and optimal control, and we investigate the extent to which these works lend themselves to classical robust control. This section is largely intended to be historical, with current research in the field left to Sec.~\ref{sec:current}. Readers unfamiliar with quantum mechanics are directed towards one of the many excellent textbooks on the subject, including Refs.~\cite{griffiths} for a popular undergraduate-level text and~\cite{nielsen_chuang,sakurai} for more advanced texts.

Although, as shown in Sec.~\ref{sec:basic}, classical control is not readily applicable to quantum systems, there are some exceptions. Petersen~\cite[Sec.~2.3.4]{Petersen2010},~\cite{Petersen2008,Petersen2013} developed quantum optical systems modeled as \emph{non-commutative} linear quantum stochastic differential equations (QSDE) and showed that such systems easily lend themselves to control by classical $H^\infty$ techniques~\cite{PAJ91}. In particular, a fundamental limitation on the disturbance rejection has been proposed~\cite[Sec. 4.2.4]{Petersen2010}. However, other quantum control designs do not allow an easy solution with the aforementioned techniques. One example is the use of time-invariant (static) but spatially distributed~\cite{Tarn_SICON} controls with structured uncertainties to maximize fidelity for, e.g., networks of quantum spins~\cite{Schirmer2018}; this type of control is referred to as robust \emph{energy landscape control}~\cite{rabitz_2007}.

As discussed, the main difficulty with quantum robust control is that in the coherent case, the closed-loop system remains purely oscillatory~\cite{soneil_mu}, invalidating all classical robust control designs that have closed-loop stability at their core. Even with the stabilization provided by decoherence~\cite{CDC_decoherence}, the Bloch equations still have a pole at zero, a manifestation of the constancy of the trace of the density matrix.

Wiseman and Milburn~\cite{wiseman_milburn} introduce quantum feedback via homodyne detection, which is very close to the classical control paradigm of feeding back the signal to be controlled. However, with the physical parameters entering the linear state space equations in a non-linear fashion, this calls for some non-trivial extension of robust control subject to structured uncertainties~\cite{schirmer2021robustness}; yet, there is a class of open quantum systems that can be viewed as the quantum Heisenberg picture analog of linear time-invariant stochastic systems, as discussed in~\cite{quantum_feedback}, which provides a springboard to Linear Quadratic Gaussian (LQG) design~\cite{Petersen2010} including Kalman filtering.

In the area of mostly open-loop quantum optimal control, there have been many works~\cite{doherty2000quantum,quantum_feedback,rabitz2004quantum} with some examining optimal control under uncertainty~\cite{rabitz2002optimal}. For example, Dahleh~\cite{dahleh1990optimal} examined the usefulness of the cost-averaging technique in solving the quantum optimal control problem subject to uncertainties. In ~\cite{zhang1994robust}, the authors formulated the robust worst-case control of quantum molecular systems using a minimax formulation and provided conditions for solving this robust control problem under certain constraints on the perturbation, which we review in Section~\ref{subsec:rcf}. In this context, the quantum control landscape, a mapping between time-dependent controls and their associated values of the objective function, has been studied to examine the analytical and numerical solutions to explore the landscapes~\cite{rabitz_2007}. The survey~\cite{Rabitz2010} covers quantum control from a historical perspective starting with magnetic resonance control, then providing an overview of the quantum control landscape, and ending with modern LQG-type control applications. A more recent survey~\cite{koch2022} covers controllability, control techniques, and applications in quantum technologies from a primarily European perspective.

\section{Linear Robust Control and Quantum Systems}\label{sec:basic}

Linear robust control is restricted to controlling a subclass of continuous-time systems known as linear time-invariant (LTI) systems to achieve a desired performance, which includes stabilization, regulation, and tracking. In this section, we explain the difficulties in applying linear robust control to quantum dynamics and why new research is needed in this area.

\subsection{State-Space Representation of Quantum Dynamics}

In control theory, the system is often represented in terms of its input, output, and state, where the state of the system depends on the particular system being modeled. In the context of quantum systems, the quantum state is represented by a vector $\ket{\psi}$ (here, represented in typical ket notation) or a density matrix $\rho:=\ket{\psi}\bra{\psi}$. In general, state vectors considered in control theory are real, but in quantum control, these are usually complex-valued. In what follows, we consider complex-valued systems, although we can easily construct real representations of the quantum state vector or density matrix via the Bloch representation, which is described in detail in Refs.~\cite{bloch1,bloch2,CDC2022}.

The state-space representation of a linear dynamical system in control theory is
\begin{equation}\label{eq:time-varying_LTI}
  \dot{\vec{x}}(t)  = \mathbf{A}_0(t)\vec{x}(t)+\mathbf{B}(t)\vec{u}(t), 
\end{equation}
where $\vec{x}(t)\in\mathbb{C}^{n}$ is the \emph{state} of the system, $\vec{u}(t)\in\mathbb{C}^{m}$ is the \emph{input} (or \emph{control}) that is designed to achieve particular specifications. The time-varying matrices $\mathbf{A}_0(t)\in\mathbb{C}^{n\times n}$ and $\mathbf{B}(t)\in\mathbb{C}^{n\times m}$ map the effect of the states and inputs on the evolution of the states, respectively.  In classical control theory, all elements in Eq.~\eqref{eq:time-varying_LTI} are real, but this is not necessarily the case for quantum systems.

Before the system is controlled, the unforced system dynamics are examined, which are written as
\begin{equation}\label{eq:unforced_time-varying_LTI}
  \dot{\vec{x}}(t) = \mathbf{A}_0(t)\vec{x}(t).
\end{equation}
The time-varying matrix $\mathbf{A}_0(t)\in\mathbb{C}^{n\times n}$ maps the state to its evolution, analogous to the time-dependent Schr{\"o}dinger equation describing the evolution of closed quantum systems. That is, in the context of control theory, a generic quantum system is represented in the state-space by
\begin{equation}\label{eq:Schroedinger}
  \dot{\vec{x}}(t) = -\frac{\imath}{\hbar}\mathbf{H}_0(t)\vec{x}(t),
\end{equation}
where the time-varying $\mathbf{H}_0(t)\in\mathbb{C}^{n\times n}$ is the same as the $\mathbf{A}_0(t)\in\mathbb{C}^{n\times n}$ matrix in Eq.~\eqref{eq:unforced_time-varying_LTI}, modulo multiplication by $-\imath/\hbar$, where $\hbar$ is Planck's constant. Eq.~\eqref{eq:Schroedinger} is indeed nothing more than the aforementioned Schr{\"o}dinger equation. Here, the state is not represented by a ket $\ket{\psi}$, but rather by a vector $\vec{x}$. The two forms of notation are equivalent. 

In the case of unitary evolution with a density matrix $\rho=\ket{\psi}\bra{\psi}$, the Schr{\"o}dinger equation takes the form
\begin{equation}\label{eq:vonN}
  \imath\hbar \frac{d}{d t} \rho = [\H_0,\rho],
\end{equation}
which is known as the Liouville-von-Neumann equation and is equivalent to the Schr{\"o}dinger equation in the case of pure states. Like the Schr\"odinger equation, the Liouville-von-Neumann equation can be rewritten in state space format by defining an orthonormal basis $\set{\sigma_\ell}_{\ell=1}^{n^2}$ for the $n \times n$ Hermitian matrices. Relative to that basis, $\rho$ is represented by the state-space vector $\vec{r}$, whose $\ell^{th}$ component is defined as 
\begin{equation}\label{eq:bloch_4}
  r_\ell(t) = \Tr(\sigma_\ell \rho(t)). 
\end{equation}
The nominal state equation of Eq.~\eqref{eq:vonN} then takes the form 
\begin{equation}\label{eq:bloch_1}
  \dot{\vec{r}}(t) = \mathbf{A}_0 \vec{r}(t), \mbox{where   } \{\mathbf{A}_0\}_{\ell,p} = \Tr(i(\mathbf{H}_0)[\sigma_\ell,\sigma_p]), 
\end{equation}
with nominal solution $\vec{r}(t) = e^{\mathbf{A_0}t}\vec{r}_0$. The resulting $n^2$ vector and $n^2 \times n^2$ matrix $\mathbf{A}_0$ are all real. This is the so-called \emph{Bloch representation}~\cite[Sec. II]{bloch2},~\cite[Sec. VII]{oneil_ls}.

\subsection{Difficulties in Applying Linear Robust Control to Hermitian Quantum Systems}\label{subsec:difficulties}

To control a generic quantum system, a control input $u(t)\in\mathbb{R}^m$ must be introduced. 
As such, the Hamiltonian of the original system changes with the addition of the input. The new \emph{nominal} dynamics are written as
\begin{equation}\label{quantum_sys_hamiltonian}
  \dot{\vec{x}}(t) = -\frac{\imath}{\hbar}[\mathbf{H}_0+\mathbf{H}_{c}(t)]\vec{x}(t),
\end{equation}
where $\mathbf{H}_c(t)\in\mathbb{C}^{n\times n}$ incorporates the control action as $\mathbf{H}_c(t)= \sum_{\ell=1}^{m} \mathbf{H}_{c_{\ell}} f_{\ell}(t)$, where $f_\ell(t)\in\mathbb{R}$ is the strength of the control field~\cite{RobustQCtrl}. The resulting system, however, is no longer strictly linear; it is bilinear since the control strengths $f_\ell(t)$ are being multiplied by the state $\vec{x}(t)\in\mathbb{C}^{n\times 1}$. 

Note that the classical control formulation of Eq.~\eqref{eq:time-varying_LTI} is easily recovered from Eq.~\eqref{quantum_sys_hamiltonian} by setting
\begin{equation}\label{eq:fictitious}
  \vec{u}(\vec{x},t) = \sum\limits_{\ell=1}^{m} \mathbf{H}_{c_{\ell}} f_{\ell}(t) \vec{x}(t),
\end{equation}
along with $\mathbf{A}_0 = -\imath \mathbf{H}_0/\hbar\in\mathbb{C}^{n\times n}$, $\mathbf{B} = -\imath I/\hbar$. 

As closed quantum systems (i.e. those that evolve unitarily) are bilinear, linear robust control generally cannot be directly applied to quantum dynamics. However, this limitation can be circumvented by considering time-invariant control strengths $f_\ell$, and defining a multivariate \emph{fictitious} state feedback $\vec{u}(t)$ as in Eq.~\eqref{eq:fictitious} with the time dependency of $f_\ell$ removed.

The concept of fictitious state feedback could not have been better articulated than by Nijmeijer~\cite{bilinear_constant_input}:
\begin{quote}
    \emph{``\dots{} the feedback is extremely simple; using the parameter choice $u=\alpha_0$} [Eq.~\eqref{eq:fictitious} with $f_\ell$ constant] \emph{does not require the knowledge of the state of the system and is therefore easy to implement. Formally, we should not call $u=\alpha_0$ a feedback\dots{} but we will use this terminology to emphasize the relation to feedback stabilization.''}
\end{quote}

\begin{figure*}[!t]
    \centering
    \scalebox{0.6}{\includegraphics{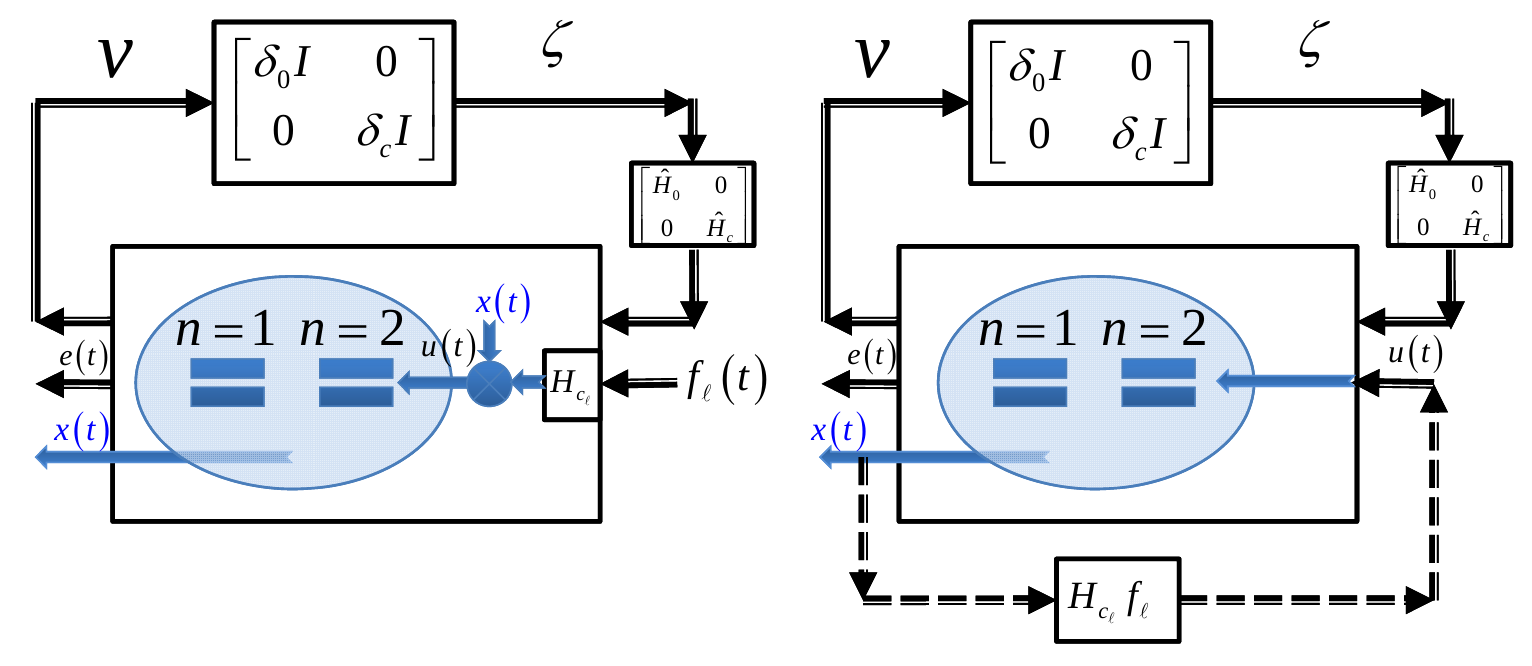}}
    \caption{Pictorial display of a bilinear \emph{open-loop} control that can be redrawn as a \emph{closed-loop} control system with expected robustness~\cite{Kosut_Phy_Rev,bilinear_constant_input}. The dashed arrows indicate that the feedback is ``hidden.'' While the illustration is specific to entanglement generation between two qubits in a cavity~\cite{RobustQCtrl}, the message is meant to be generic, aimed at any quantum control problem insofar as it is bilinear. The state could be the wave function $x(t)$ as in Eq.~\eqref{eq:fictitious}, the density operator $\rho(t)$, or the Bloch vector $r(t)$. The error signal $e(t)$ is, here, the concurrence error $1-C_{ss}$ as in Sec.~\ref{sec:with_mu}, although it could also be the fidelity error $1-F$ as in Eq.~\eqref{eq:vectoized_fidelity_error}. The structure of the uncertainties $\hat{\H}_0, \hat{\H}_c$, and their strengths $\delta_0, \delta_c$ (see Eq.~\eqref{eq:uncertainty}) are expressed as feedback around the plant as is traditionally done with structured perturbation~\cite{Zhou}.}
    \label{fig:the_pitch_of_the_paper}
\end{figure*}

A valid counterargument is that the same scheme could be interpreted as open-loop control, hence without robustness properties to be expected. However, as argued by Kosut~\cite{Kosut_Phy_Rev}, the control input to a bilinear system is mixed with the state and therefore the hidden feedback effect could result in resistance against uncertainties. This concept is illustrated in Fig.~\ref{fig:the_pitch_of_the_paper}.

As a brief example: the single spin-$\tfrac{1}{2}$ particle can be written as an LTI system, as in Eq.~\eqref{quantum_sys_hamiltonian}, in the following manner. Let $\mathbf{H}_0 = \frac{\omega_q}{2}\sigma_z$, $\mathbf{H}_{c_{1}} = \sigma_x$, $f_1(t) = u_x$, $\mathbf{H}_{c_{2}}=\sigma_y$, and $f_2(t) = u_y$, where $\omega_q$ describes the bare qubit resonance frequency, and the controls $u_x$ and $u_y$ describe motion about the $x-$ and $y-$axes of the three-dimensional Bloch sphere. Here, the $\sigma_\ell$ represent the well-known Pauli matrices for $\ell = x, y, z$; multi-qubit generalizations of these matrices are known as the Gell-Mann matrices~\cite{Bertlmann2008}. Furthermore, to obtain time-invariance, one must restrict the control inputs to be constant for all time (i.e., $u_x(t) = u_x$ and $u_y(t) = u_y$). Otherwise, the system is not describable as an LTI system. Note that both closed and open quantum systems can be written as LTI systems, and we present here a closed system for simplicity.

\subsection{Stability of Quantum Systems}

In the case of Hermitian (closed) quantum evolution, the eigenvalues of the system are purely imaginary. This is because the eigenvalues of the Hermitian matrix $\H=\H_0+\H_c$ are purely real, thus those of 
$\A=-\imath \H/\hbar$ are purely imaginary. Despite this lack of stability by classical standards, such oscillatory systems can still be evaluated relative to some performance metric, typically fidelity at a terminal time $t_f$. 
Application of classical methods to such Hermitian systems would require the addition of some dissipative, non-Hermitian quantities to the dynamics. This option is given quite naturally by the so-called \emph{open} quantum systems, where state evolution is not strictly unitary. The control of both open and closed systems is discussed in Sec.~\ref{sec:struct}.

In contrast to closed quantum systems governed by the Schr{\"o}dinger equation, the evolution of open quantum systems takes the form of the Lindblad master equation~\cite{Manzano}, an open system extension of Eq.~\eqref{eq:vonN}, 
\begin{equation}\label{eq:L_master}
  \frac{d}{dt} \rho = -\frac{\imath}{\hbar} [\H,\rho] + \sum_k\gamma_k\left(V_k \rho V_k^\dagger - \frac{1}{2}\{V_k^\dagger V_k,\rho\}\right),
\end{equation}
where the anticommutator of two operators is defined as $\{A,B\} = AB + BA$. The nonunitary behavior is seen in the second term, where the Lindblad jump operators $V_k$ define the type of dissipation present in the system and $\gamma_k$ their strengths (called damping rates). The Bloch representation is still applicable~\cite{neat_formula}; indeed, along with Eq.~\eqref{eq:bloch_1}, it suffices to add
\[\{A_L\}_{\ell,p}=\sum_k \left(\Tr(V_k^\dagger\sigma_{\ell} V_k \sigma_p)-\frac{1}{2}\Tr(V_k^\dagger V_k \{\sigma_\ell, \sigma_p\})\right).\]
This approach is somewhat deceptive in the sense that it does not reveal that the decoherence depends on the system Hamiltonian and hence on the control, especially when the control fields are time-varying~\cite{domenico_CDC,singular_vs_weak_coupling}. One way to force the Lindblad model to deal with this dependency is to impose decoherence to act in the Hamiltonian basis~\cite{RobustQCtrl}, that is, 
$[V_k,\H]=0$, although the challenge then is finding methods of imposing this that still allow for an accurate description of the system dynamics.

Despite its often undesirable omnipresence, dissipation takes the problem of applying classical methods to quantum systems towards potentially new paradigms. In the aforementioned case of unitary evolution, where all the system's poles lie on the imaginary axis~\cite{CDC2022}, the system has zero stability margin. If the system is modified to include non-unitary behavior (like decoherence and decay) in the system, the resulting LTI system is modified such that its poles have negative real parts (after removing the pole at zero reflecting the fact that $\Tr{(\rho)} = 1$~\cite{RobustQCtrl}). Armed with this formalism, the classical notions of stability and robustness can be applied to an open quantum system~\cite{RobustQCtrl}. As with classical control metrics, one can also analyze the steady-state behavior of the system. That is, one can detect whether there is a unique steady state that is globally attractive or there exists a continuum of steady states. There is, however, a tradeoff. Classical control is typically interested in the long-time behavior of a system, including the steady states of the system. However, with some exceptions (see, e.g.,~\cite{Motzoi2016}), the steady states of open quantum systems are not interesting as they offer no quantum advantage. Hence, quantum practitioners are typically interested in exploring short-time dynamics, such as optimizing quantum gates to be as fast as possible. As such, there is interest in the development of robust quantum control that can be applied in the time domain, which is discussed next in Sec.~\ref{sec:struct}.

\section{Robust Quantum Control in Practice}\label{sec:struct}

This section of the survey deals with applying robust control in practice. Finding a general theory of quantum robustness is an active field of research. There is a practical need for analytic methods for determining robustness that do not require extensive sampling. One can think of this as the quantum analog to the well-developed theory of classical control. However, due to the limitations discussed earlier, developing such a theory in the quantum regime is not straightforward. This is yet another example of an arena where quantum robust control is lagging: while classical robust control theory can determine the maximum possible perturbations that can be applied to a given system, regardless of the size of the perturbations, a general technique for applying this to quantum systems is not readily available. Thus, the intention here is to describe the methods that are commonly used in the field, with relevant citations for the more deeply interested reader. We provide a practical survey of techniques in the hope that better methods will be found in time.

It is important to note the difference between error correction and robustness. Error correction actively seeks to find errors in, e.g., the state of a quantum computer. This can be done in a myriad of ways, the simplest of which is a rudimentary repetition code that runs the same sequence multiple times and uses this to discard erroneous measurements. Robustness, on the other hand, seeks to find means to control a system (with so-called ``controllers'') in such a way that the controller itself is robust to uncertainties, which may arise due to noise, decoherence, and other inevitable systematic errors that may occur in an experiment. Thus, a truly robust protocol with sufficiently high fidelity would not require any error correction. Whether or not such protocols can exist in practical quantum technologies, like quantum computing, is an open question, which lies outside the scope of this survey.  Likely, some error correction will still be needed even with the most robust controls, but this is an important question for future research.  

In what follows, we first describe how to model uncertainty in quantum systems by introducing the concepts of fidelity, as well as structured and unstructured perturbations, which are important for quantifying robustness. This leads to a discussion of how to formulate the problem of robustly controlling quantum dynamics, which is explored for both closed and open quantum systems. We discuss how to find controllers for systems possessing a model of the dynamics as well as for systems where there is no underlying model, and briefly describe methods used to find good controllers in experiments. Then, we move on to the evaluation of the resulting controllers in terms of robustness. Finally, we discuss how to rank controllers by their robustness to determine the best controller for a given job.

\subsection{Robust Control Formulation\label{subsec:rcf}}

\subsubsection{Closed Quantum Systems}\label{sec:with_Pont}

Minimizing the tracking error so that the desired final state matches the actual state is similar to maximizing the fidelity between two states in quantum systems. For closed quantum systems, the fidelity is represented by
\begin{equation}\label{eq:fidelity_coherent}
  F = |\bra{\psi_{d}}U(H,t_f)\ket{\psi_{t_0}}|^2,
\end{equation}
where $U(H, t_f)$ is the unitary operator evolving the state from a time $t_0$ to $t_f$ under the Hamiltonian $H$. There exist extensions of this measure to density matrices~\cite{Koch_2016}. One can also determine the fidelity of a process, which finds utility in, e.g., quantum gate optimizations~\cite{biercuk}. It would be prudent to design a control that manipulates a system to ensure not only that the actual final state of the system is close to the desired final state, but in addition, that the fidelity $F$ does not depend too much on uncertain parameters such as the spin-spin couplings in a spin network, bias fields~\cite{rabitz2002optimal,Koswara_2021}, the shape of the control pulses~\cite{Gneiting}, etc. This is fundamentally what robust quantum control aims to achieve although the notion of what ``too much'' actually entails in practice is often system- and problem-dependent.

Robust quantum control probably has its inception in the early paper by Rabitz~\cite{zhang1994robust} on quantum chemistry and the work of Khaneja~\cite{grape} in nuclear magnetic resonance. It was later taken on in the context of quantum optics by Petersen~\cite{Petersen2013}. Over the past few years it has progressed at an accelerated pace~\cite{biercuk,Koswara_2021,Gneiting,Guerin2020,Wu2021,Kestner,Irtaza_PRA,kosutrabitz,Yung2021,Mahesh,Hensinger,Glaser2021,Wang2019} in quantum error suppression, entanglement control, cross-talk control, quantum computers, etc. However, in this survey, we focus on state transfer with a view to modern robust control. To that effect, we model uncertainty in a way amenable to structured perturbations; that is, referring to Eq.~\eqref{quantum_sys_hamiltonian}, 
\begin{equation}\label{eq:uncertainty}
\tilde{\H}_0+\tilde{\H}_c=(\H_0+\delta_0 \hat{\H}_0) + (\H_c+\delta_c \hat{\H}_c), 
\end{equation}
where the tildes, $\tilde{\H}_0$, $\tilde{\H}_c$, denote perturbed quantities, whereas the hat notations, $\hat{\H}_0$, $\hat{\H}_c$, denote the properly normalized \emph{structures} of the perturbations, and $\delta_0$, $\delta_c$ denote the \emph{strengths} of the perturbations on the drift dynamics and the control couplings, respectively. 

A first approach consists in maximizing $F$ or minimizing the ``tracking error'' $1-F$, and then analyzing the sensitivity/robustness of the design. Coherent quantum systems appear to circumnavigate the classical limitations and allow for small error $1-F$ to coexist with small logarithmic sensitivity~\cite{statistical_control}. However, such behavior, which contradicts fundamental limitations on achievable control performance~\cite{Bode1945,Safonov_Laub_Hartmann}, is not universal and disappears under decoherence~\cite{Cambridge_decoherent,Cambridge_coherent}. The latter is not surprising, since decoherent systems are ``closed-loop stable'' and, hence, more conventional.

The second approach is about robust \emph{design} rather than \emph{analysis} and aims at finding the optimal controller for a weighted combination of the final fidelity error and its sensitivity, in a design that has its roots in the \emph{mixed-sensitivity} $H^\infty$ design~\cite{mixed_achievable_performance}:
\begin{equation}\label{eq:2_criteria} 
  \min_u \left( \alpha (1-F(t_f)) + \beta \|\nabla_\delta F(t_f)\|^2 \right),
\end{equation}
where $\nabla_\delta$ is the vector of partial derivatives of $F$ relative to $\delta_0, \delta_c$ evaluated at $\delta_0=\delta_c=0$. This is basically the infinitesimally small perturbation approach dating back to Bode~\cite{Bode1945}.  However, in classical control a dramatic shift towards \emph{large} perturbations came about~\cite{Safonov_Laub_Hartmann} in the 1980s. The large perturbation approach is embodied in the $\mu$-function or \emph{structured singular value}, among other techniques. However, the application of the $H^\infty$ control techniques and structured singular value analysis, in particular to the robustness of quantum control problems, is fraught with difficulties~\cite{jonckheere2017structured} (explained in the simplest possible setting in~\cite{schirmer2021robustness} and reiterated in Sec.~\ref{sec:with_mu}). Its application in a quantum context is very limited, mostly restricted to dissipative quantum systems for the simple reason that $H^\infty$ robust control techniques mainly aim to synthesize controllers that achieve stabilization with guaranteed performance, and, as outlined previously, for most quantum control problems, stabilization is undesirable.

Most of the approaches investigating fidelity error versus sensitivity simplify the problem to be finite-dimensional, i.e., the controller is a finite set of static bias fields, piecewise time-invariant bias fields, or a train of pulses.  An alternative approach is based on the Pontryagin maximum principle.  This approach allows the control $u(t)$ to be continuous in time and allows the designer to force the final fidelity error to be $1-F(t_f)=0$, and minimize a weighted combination of $\|\nabla_\delta F(t_f)\|$ and $t_f$ relative to $u(t)$.  In the spirit of achieving the desired final state in the minimum amount of time, the criterion becomes 
\begin{equation}\label{eq:Pontryagin} 
  \min_{u(t), t_f} \left( \beta\|  \nabla_\delta F(t_f)\| + \gamma t_f \right).
\end{equation} 
A quantum Pontryagin Maximum Principle was introduced in~\cite{boscain2020introduction}. Belavkin~\cite{BelavkinarXiv} introduced the competing Bellman Principle of Optimality in quantum control problems, which later resurfaced as \emph{dynamic programming}~\cite{robust_closed_loop_quantum}.  Recent work
by Sugny~\cite{garon2013time} on the pure brachistochrone problem ($\beta=0$) leads to the observation that the minimum time to reach the desired state depends on the global phase of $\psi(t_f)$.  This surprising result and some of its ramifications~\cite{bian2019experimental} suggest that this will remain an active research area~\cite{CDC2022Pontryagin}. 

\subsubsection{Open Quantum Systems}

The control of open quantum systems~\cite{TSH2012} is still very much an open research area. An excellent review (with examples) can be found in~\cite{Koch_2016}, and a two-part manuscript on the limits of open system control can be found in~\cite{TSH2,TSH1}. There are many open questions in the area of controllability of open quantum systems and the effect of non-Markovianity on control. We summarize some key points here.

Given an initial state $\rho_{t_0}$, a Hamiltonian $\H$ and a set of Lindblad operators $\{V_k\}$ defining a Lindblad master equation as in~\cite{Manzano} and Eq.~\eqref{eq:L_master}, one can often use similar methods to find and optimize controllers. An example using the Krotov method is given in~\cite{Koch_2016} and other methods are described in~\cite{Koch_2019,Kosloff_2022}.

Fidelity measures for open quantum systems must be slightly modified when dealing with density operators not representing pure quantum states. A variety of different fidelity measures are available~\cite{Liang_2019} to define a fidelity between two mixed states (i.e. states that cannot be written in bra-ket form). The most common is the Uhlmann-Jozsa fidelity, which reduces to the standard fidelity (Eq.~\eqref{eq:rho_fidelity}) if one state is pure~\cite{Koch_2016}. Given this fidelity, other terms similar to those in Eqs.~\eqref{eq:2_criteria} and~\eqref{eq:Pontryagin} can be added, taking into account the differences between pure states and density matrices.

To minimize the deleterious effects of decoherence for open systems, it is often desirable to work in a \emph{decoherence-free subspace (DFS)}~\cite{lidar} --- if such a subspace exists. In laser cooling of an atom~\cite{metcalf}, for example, one can define a set of atomic transitions that define a closed system. Typically, these are one cycling (i.e. short-lived) transition for cooling and some repumping transitions that catch atoms that decay to states other than the two levels defined by the cycling transition. All other transitions in the atom are forbidden to the extent that they are never populated, so the cycling and repumping transitions define such a decoherence-free subspace. In geometric control terms, such subspace is known as a \emph{null-output, controlled-invariant subspace} (or \emph{distribution} in the nonlinear case)~\cite{Tarn,DSM_expanded}. Control-invariance means that, with suitable controls, decoherence confines the density to a nontrivial invariant subspace (or distribution);  null-output means that any initial state in such a subspace (or distribution) creates zero transfer error. If the initial preparation error initializes the state outside the decoherence-free subspace, transients will occur, and it is still unclear how to achieve a fidelity tolerance within the shortest time $t_f$, thus minimizing the effect of deleterious processes. Open-system versions of Eqs.~\eqref{eq:2_criteria},~\eqref{eq:Pontryagin} are useful here. 

\subsubsection{Bloch Vector Formulation}\label{sec:with_mu}

To draw a closer parallel to classical robust control, one can rewrite the Hamiltonian perturbation (Eq.~\eqref{eq:uncertainty}) in accordance with the real Bloch representation of the dynamics of Eq.~\eqref{eq:vonN}, cf. Sec.~\ref{sec:basic} and Section~II of~\cite{bloch2}
\begin{subequations}
\begin{align}
  \left(\mathbf{S}_{c}\right)_{\ell p} &= \Tr(i\mathbf{\hat{H}_c}[\sigma_\ell,\sigma_p)]), \\ 
  \left(\mathbf{S}_{0}\right)_{\ell p} &= \Tr(i\mathbf{\hat{H}_0}[\sigma_\ell,\sigma_p)]).\label{eq:bloch_3}
\end{align}
\end{subequations}
The perturbed dynamics are then given by $\dot{\tilde{r}}(t)= (\mathbf{A}_0 + \delta_0 \mathbf{S_0} + \delta_c \mathbf{S_c})\tilde{r}(t)$, where $r(t)$ is the Bloch representation of the state $\rho(t)$. For constant control fields (time-invariant case), the solution is $\tilde{r}(t) = e^{(\mathbf{A} + \delta_0 \mathbf{S_0} + \delta_c \mathbf{S_c})t} r_0$.

For the state-transfer problem, if the target state $\rho_d$ is the pure state $\ket{\psi_d}$, then the nominal fidelity cf. Eq.~\eqref{eq:fidelity_coherent} can be expressed in terms of the Bloch vectors:
\begin{align}\label{eq:rho_fidelity}
\begin{split}
  F &= \Tr(\bra{\psi_d} \rho(t_f) \ket{\psi_d}) = \Tr(\rho_d \rho(t_f))\\
    &= \sum_{\ell,p = 1}^{n^2} \left(r_{d}\right)_\ell r_p(t_f) \Tr(\sigma_\ell \sigma_p) = r_d^{T}r(t_f).
\end{split}
\end{align}
If the basis set $\set{\sigma_\ell}$ is chosen such that $\sigma_{n^2} = (1/\sqrt{n})I_n$, then the requirement that $\Tr(\rho(t)) = 1$ implies that $r_{n^2}(t) = 1/\sqrt{n}$ for all $t$. Then defining $\sqrt{n}{e}_{n^2}$ as the $n^2 \times 1$ column vector with all entries zero, save for the last entry which is $\sqrt{n}$, the result is $\sqrt{n}{e}_{n^2}^T r(t) = 1$. With  $c = \sqrt{n} e_{n^2} - r_d$ one may now rewrite the nominal fidelity error as
\begin{equation}\label{eq:vectoized_fidelity_error}
  e(t_f) = 1-F = c^Tr(t_f) = c^Te^{\mathbf{A}t_f}r_0.
\end{equation}
Defining the operator $W: r_d \mapsto r_0$ as the Bloch representation of the map $\ket{\psi_d} \mapsto \ket{\psi_0}$, the above becomes $e(t_f)=c^Te^{\mathbf{A}t_f}Wr_d$. That is, $\mathcal{S}(t) := c^Te^{\mathbf{A}t_f}W$ maps the desired output to the error~\cite{statistical_control}. Since Bode~\cite{Bode1945}, this mapping has been referred to as the \emph{sensitivity function}, and questions about the sensitivity of $\mathcal{S}$ to potentially large variations in $\mathbf{A}$ essentially launched modern robust control~\cite{Safonov_Laub_Hartmann}. However, most research on the sensitivity function has been conducted in the frequency domain and the transcription into the more-quantum-relevant time domain remains challenging.

Despite a nontrivial frequency to time domain transcription, some hope for constraints involving $\mathcal{S}(t)$ and its sensitivity can be found if the variation is differential and unstructured, $d\A$, and commutes with its nominal, $[\A,d\A]=0$. Then, $\mathcal{T}(t)=c^Te^{\A t}r_0$ appears to be the sensitivity of $\mathcal{S}(t)$ to $d\A$ and leads to the limitation~\cite{Cambridge_coherent}
\[
  r_d^T(\mathcal{T}(t)\mathcal{T}(t)^T+\mathcal{S}(t)^T\mathcal{S}(t))r_d=1.
\]
In the large, structured perturbation case, $\tilde{\A}=\A+\delta_0\A_0$, the Zassenhaus formula~\cite{Zassenhaus}, which expresses the multiplicative deviation of the matrix exponential from its nominal value in terms of the Lie polynomials of $\A_0t$ and $\delta_0 S_0t$, offers some hopes for a generalization of the sensitivity limitation.

Traditionally structured perturbations are dealt with using the structured singular value or $\mu$-function~\cite{Zhou}.  Besides the frequency to time domain transcription, its application to quantum problems involves further difficulties: (i) in the Schr\"odinger` and Lindblad formalisms, there is no noise response relative to which the effect of uncertainties can be gauged; (ii) there are poles at zero and a lack of stability. The first difficulty is overcome by resorting to a noise-agnostic approach~\cite{RobustQCtrl}. The second can be circumvented in the dissipative case by a specialized ``pseudo-inverse'' to deal with the pole at $0$ that remains due to trace constraints~\cite{RobustQCtrl}. In this case, the most straightforward way to deal with structured uncertainties appears to be to consider the mapping $\tilde{T}_{t_f,0}(\delta_0): r_0 \mapsto c^T\tilde{r}(t_f)$ where only one uncertain strength $\delta_0$ is considered for simplicity. Following~\cite{schirmer2021robustness}, $\mu$ can be defined such that $\|\tilde{T}_{t_f,0}(\delta_0)\|\leq \mu$, $\forall \delta_0 \leq 1/\mu$. Under the assumption that the super-level sets $\{\delta_0: \|\tilde{T}_{t_f,0}(\delta_0)\|\geq c\}$ are convex and unbounded, it can be shown that in the definition of $\mu$ the inequalities can be replaced by equalities. Hence, $\mu=1/\delta$ where $\delta$ is given by the fixed point problem $\lVert \tilde{T}_{t_f,0}(\delta) \rVert =1/\delta$.  This approach does not require the error signal to be linear in the state and applies to nonlinear performance like the concurrence error $1-C_\text{ss}$, where $C_\text{ss}$ is a measure of entanglement~\cite{Domenico_concurrence,EntanglementConcurrence}. Regarding design, instead of Eq.~\eqref{eq:2_criteria}, one would use the single criterion~\cite{jonckheere2017structured,schirmer2021robustness}
\begin{equation}\label{mu_design} 
  \min_u \mu(G), 
  \end{equation}
where $G$ is a $2\times 2$ partitioned matrix which, with the uncertainty as a feedback wrapped around it as shown in Fig.~\ref{fig:the_pitch_of_the_paper}, reproduces $\tilde{T}_{t_f,0}(\delta_0)$.  Eq.~\eqref{mu_design} has the advantage of combining in a single criterion fidelity error and robustness for \emph{large} perturbations. However, as stated previously, this approach is only applicable to open systems where decoherence provides stability margins and singularities can be removed under certain conditions. Moreover, for many quantum control applications, stabilization is generally undesirable.

\subsection{Finding Controllers}

As already introduced in Sec.~\ref{sec:with_Pont}, the Pontryagin Maximum Principle~\cite{boscain2020introduction,CDC2022Pontryagin} and the Bellman Principle of Optimality or dynamic programming~\cite{robust_closed_loop_quantum} are insightful analytical techniques~\cite{garon2013time} that endeavor to find an optimal controller \emph{analytically}. However, more research is needed to apply these techniques to solve the robust control problems arising for quantum systems such as Eqs.~\eqref{eq:2_criteria} and~\eqref{eq:Pontryagin}. Furthermore, the relevant target functions are non-convex, and optimization of such problems is non-trivial~\cite{bertsekas2000dynamic}.

Another analytical approach considers \emph{shortcuts to adiabaticity} in quantum control. These methods make use of inverse engineering methods, counter-diabatic driving, and other ``shortcut'' methods to reproduce adiabatic behavior in quantum systems without the long timescales required to strictly maintain the adiabaticity condition. These methods have a wide range of applicability within different quantum fields (e.g. state transfer~\cite{STIRAP} or particle transport~\cite{STA_transport}). In many cases, a degree of robustness is built into the system design, e.g., by requiring 
\begin{equation}\label{eq:STA}
  \dot{u}(t = 0) = \ddot{u}(t = 0) = \dot{u}(t_f) = \ddot{u}(t_f) = 0
\end{equation}
for a time-dependent control $u(t)$ running from time $t = 0$ to $t_f$. An extensive review of such methods is outside the scope of this survey, but an excellent recent review can be found in~\cite{STA}.

In cases where analytic methods become intractable, optimization algorithms are useful. The class of optimizer that one chooses depends on the problem at hand. In this section, we describe the methods most commonly used in the physics community to differentiate the relevant use cases for each class of method. We also focus on the differences in approach between theoretical and experimental models.

In terms of optimization algorithms, one can differentiate common methods in terms of whether or not they rely on gradients in the control landscape. Broadly, for a control $\vec{u}$, one can iteratively update the control with respect to the gradient of a performance measure, usually the control fidelity $F$, as
\begin{equation}\label{eq:gradientF}
  \vec{u} \rightarrow \vec{u} + \alpha\nabla F\{\vec{u}\}, 
\end{equation}
where in some cases, the parameter $\alpha$ is tunable as the protocol converges. Common gradient-based optimizers include Krotov's method~\cite{krotov}, stochastic gradient ascent methods~\cite{QM_SGA,Sels_SGA}, and the GRAPE method~\cite{grape}. The GRAPE method has recently been extended to a risk-sensitive version (RS-GRAPE) that can optimize for robustness as well~\cite{Wu2021}; likewise, the Krotov method has been extended for the optimization of controllers in noisy environments~\cite{Gneiting}. Quasi-Newton methods like the L-BFGS method~\cite{lbfgs} are also commonly used, and modern versions of GRAPE are also typically quasi-Newton in nature. In a theoretical setting, when one has access to a \emph{model} of the system at hand (i.e., a Hamiltonian describing its evolution), typically one uses gradient-based methods where analytical derivatives are calculated explicitly or automatic differentiation may be used~\cite{autodiff}. Other recent work has explored the utility of approximate derivatives in optimal control~\cite{mohr}. The local nature of a gradient-based optimizer (in that it simply climbs the nearest hill and ignores all other optima in the landscape) is overcome by \emph{restarting} the algorithm multiple times with different initial conditions, or \emph{seeds}~\cite{multistart}.

Gradient-free methods do not require the calculation of a gradient, which is useful in cases where such computations are expensive or there is no available model (i.e., in a real experiment, where one only has access to the experimental inputs and outputs, but the actual Hamiltonian underlying the system evolution may be uncertain). Such methods include evolutionary methods~\cite{rabitz_ga,sanders_ev} and direct-search methods like Nelder-Mead-based methods or the related CRAB algorithm~\cite{CRAB}; the latter of these methods relies on an intelligent parameterization of a time-dependent control. For example, in lieu of Eq.~\eqref{eq:fictitious}, a time-dependent control $\vec{u}(t)$ can be parameterized as
\begin{equation}
  \vec{u}(\vec{x},t) = \sum\limits_{\ell=1}^{m} \mathbf{H}_{c_{\ell}} \left(\sum_n a_{n,\ell} f_{n}(t)\right) \vec{x}(t),
\end{equation}
for some parameters $\{a_{n,\ell}\}$ and basis set of functions $\{f_n(t)\}$, e.g., a Fourier, wavelet, or Chebyshev basis. There exist methods that combine gradient-based search with the parameterization of the CRAB methods, e.g., the GROUP method~\cite{GROUP}. As stated previously, in an experimental setting, gradients are often expensive to evaluate, especially if the experimental repetition rate is low. Although in experiments with high repetition rates gradient-based algorithms have been successfully used~\cite{rabitz_lab}, the required numerical derivative calculations are typically sub-optimal relative to the use of analytically-derived gradients.

In addition, it is often desirable to parameterize a time-dependent control or constrain the controls to narrow the available search space or to better model experimental settings, e.g., by filtering a time-dependent control to remove high-frequency components or applying the basis constraints in the aforementioned CRAB algorithm. Constraints are often implemented to model limitations in the control field amplitude and bandwidth that are present in real systems. Note that this parameterization is different from the parameterization imposed by the numerical need to discretize time: a control parameterized, e.g., by a set of Fourier components can be developed that maintains the same time discretization as an unparameterized control. These restrictions often lead to local \emph{traps}, or local optima, within a search landscape, but this can be overcome by employing a restart strategy with careful sampling of initial values. The idea here is that with a good parameterization of the controls that relies on a prior understanding of the physics of the problem (e.g., limiting a search in Fourier space to the relevant transition frequencies of the quantum system under consideration), one can more efficiently search the lower-dimensional landscape relative to the less-constrained but more complex higher-dimensional landscape. Controls can also be \emph{dressed} by modifying them once they have reached a local minimum moving them out of the minimum, and allowing the optimization to proceed further~\cite{dCRAB}.

Another alternative is the use of data-driven methods. Gaussian process optimization~\cite{gaussian_processes} is an example of a model-free method that is the best method for direct optimization of cold atom experiments~\cite{footGP}. The idea is to find a maximum of a function, such as the fidelity, that is expensive to evaluate (e.g. because it requires experimental results) by representing it as a probability distribution. One can gain information about the function by applying a sampling strategy to collect observations and based on the observations estimate the location of a likely maximum. The sampling strategy obtains further data points to increase the certainty of the location of the maximum.  

Reinforcement learning (RL)~\cite{rlforoptimalcontrol,Sels2018} or supervised learning~\cite{Wang2019} methods have also been employed. RL explores an environment and optimizes a policy to maximize the reward obtained by executing actions in the environment. The optimal policy determines the best action to execute to maximize the expected reward. The reward represents the problem and the environment is modeled as a Markov decision process. RL methods can be applied in both \emph{model-free} and \emph{model-based} cases. A model-free approach does not explicitly use the transition probability distribution and the reward function (together referred to as the model) associated with the Markov decision process. It resembles a trial-and-error process where the decision of the next action to execute is based on observations about the current state (the policy is learned directly). A model-based RL approach, instead, learns the transition probabilities and reward in the Markov decision process explicitly to model the environment and the policy aims to maximize the expected reward predicted by this model.

Model-based~\cite{universalQC_RL} and model-free~\cite{modelfreeRL} methods have been used to implement quantum gates and circuits. It remains an open research question whether the robustness of controllers found depends on the optimizer.  Although comparisons of algorithm efficacy have been made~\cite{irtaza,Irtaza_PRA,rl_nature}, it remains to be seen if a given RL method, for example, finds a robust controller more often than another method.

\subsection{Determining Robustness in Practice}\label{subsec:robustness_practice}

As researchers work to find general methods of performing robust quantum control, there are many ways to test notions of robustness on objective functions such as Eqs.~\eqref{eq:2_criteria}, \eqref{eq:Pontryagin}, and~\eqref{mu_design} in practice. The basic idea is the following: given a controller with reasonable performance (e.g., in terms of state transfer fidelity or implementation of a target unitary), the robustness of the controller is evaluated. One possible evaluation method is via the application of structured or unstructured perturbations. Effectively, one takes the controller, perturbs it, and quantifies how the results change. In the following sections, we discuss some methodologies used in practice to test robustness.  This is not exhaustive but intended to provide a good starting point for the robust control practitioner.

\subsubsection{The Log-Sensitivity}

For the model-based formulation, one can assess the sensitivity of the system to perturbations structured as Eq.~\eqref{eq:uncertainty} by analyzing how the transfer fidelity changes as a function of the perturbation amplitude via derivatives of the form
\begin{equation}\label{eq:deriv}
  \partial_cF:=\partial F(\tilde{\H}_0+\tilde{\H}_c)/\partial \delta_c.
\end{equation}
Such perturbations are typically more realistically evaluated if one considers the logarithmic sensitivity
\begin{equation}\label{eq:logsens}
  \partial_c\log{(F)} = \partial_c F/F,
\end{equation}
where the $\partial_c$ here implies that perturbations on the controller are perturbing the function $F$, and in many cases, these derivatives must be computed numerically instead of analytically. Computing the log-sensitivity can be somewhat expensive, especially in a high-dimensional space. However, when feasible, the log-sensitivity allows one to determine which parameters are the most and least sensitive to perturbations. This method has recently been adapted to the analysis of controllers in the time-domain~\cite{oneil_ls}, which is applicable to both closed and open quantum systems.

\subsubsection{Monte-Carlo Sampling}

It is common in quantum research to determine a controller's robustness by perturbing the controller with some probabilistic noise spectrum and sampling the system's response.  Effectively, these methods, known as Monte-Carlo or quasi-Monte-Carlo sampling~\cite{quasi_monte_carlo_robust,mc_and_qmc,qmc_quantum}, assume a probability distribution for the system noise. Sampling from this noise probability distribution could be done either randomly from a uniform distribution (Monte-Carlo sampling) or in a manner that more evenly covers the sample space (quasi-Monte-Carlo sampling). Typically, quasi-Monte-Carlo approach samples using a low-discrepancy sequence to better sample the space of noisy controllers. Such methods, which are a workhorse in quantum control, allow one to determine a distribution of fidelities given a noise distribution.  However, these methods can be expensive, in that many ($>1,000$) samples are typically needed to obtain good statistics on the fidelity distribution. Other methods based on incorporating the propagation of quantum uncertainties directly into the model have been demonstrated~\cite{Motzoi2022}; such methods can be faster than Monte-Carlo sampling and may allow physicists to move beyond Monte-Carlo methods.

\subsubsection{Ranking Controllers}

Given the controller sensitivity, determined by computation of the log-sensitivity, via Monte-Carlo sampling methods, or from experimental data, one must determine the best controller for their application.  Methods of ranking both controller fidelity robustness and control algorithm efficacy based on the Wasserstein distance between the ideal distribution (a Dirac delta at $1$) and the distribution achieved by the controller have recently been demonstrated for quantum systems~\cite{Irtaza_PRA}, and robustness measures based on the method of averaging have also been recently shown~\cite{kosutrabitz}.  In many cases, this depends very heavily on the specifics of the scenario in which the controller is deployed. For example, in some systems, one can tolerate a fidelity minimum $F_\mathrm{min}$, but the controller should be robust enough that the probability of a fidelity lower than $F_\mathrm{min}$ is vanishingly small. In other scenarios, the constraints on $F_\mathrm{min}$ may be more stringent, but the system can tolerate a nonzero probability of a lower fidelity, e.g., due to differences in how the data is treated in post-processing. Therefore, it is important, when choosing the right controller for a given application, to have a reasonable probabilistic model for the noise sources, sufficient sampling to test the noise distribution thoroughly, and an idea of the application space in which the controller will be deployed.

\section{Current and Future Research Directions}\label{sec:current}

So far in this survey we have covered certain relevant aspects of robust quantum control from a primarily (but not completely) historical perspective. In this section, we will look at many of these concepts, but now from an eye toward more current research in the field, as well as ideas for future developments in quantum robust control. In addition, we will cover relevant aspects of the robust control of experimental quantum technologies.

Concerning \emph{measures of robust performance} and \emph{performance guarantees}, classical techniques such as structured singular value theory are useful for certain applications involving open systems such as reservoir engineering~\cite{RobustQCtrl}. Indeed, there is a growing body of work on engineering system-bath interactions and measurements to realize control~\cite{Kapit2017,Hacohen-Gourgy2018,Hacohen-Gourgy2020,Harrington2018}. However, for the reasons discussed in the previous section, these techniques are of limited use for applications such as quantum gate engineering and quantum state preparation in closed systems. New ideas are needed to define what constitutes robust performance in different quantum control settings, how it should be quantified, and what performance guarantees can be given. The differential and logarithmic sensitivity are useful tools to assess robustness for small perturbation~\cite{O_Neil_2023a,O_Neil_2023b,LCSS2023}, which can provide limited performance guarantees even for time-dependent controls~\cite{O_Neil_2023c}. Further development of the aforementioned statistical robustness measures based on the Wasserstein distance~\cite{Irtaza_PRA} may also be fruitful.

Instead of assessing the robust performance of controllers after they have been designed by optimization or other means, a significant amount of work is focused on \emph{robust control synthesis}, i.e., finding controls that are robust by design. One approach to achieve this involves incorporating robustness criteria into the optimization targets, e.g., optimizing average gate fidelities~\cite{Irtaza_PRA}, minimizing the size of an error Hamiltonian~\cite{kosutrabitz}, optimizing the quantum Fisher information of the system~\cite{Poggi_2023}, utilizing Pareto optimization~\cite{Bhutoria_2022}, and adapting existing algorithms to handle noisy systems~\cite{Gneiting}. Some of these approaches can be applied to open quantum systems and can handle a wider range of perturbations, but most involve solving computationally expensive optimization problems, and good solutions are not guaranteed. To avoid such computational overheads, other efforts are focused on experimental implementations of optimal pulse engineering~\cite{grape,Mahesh,Koch_2019,Werninghaus_2021,Wittler2020}, and direct methods to find robust controllers in experiments~\cite{Ferrie,Weidner,PhysRevA.80.030301}, as well as hybrid methods that rely on both experiments and models~\cite{chakrabarti}. 

In the area of quantum gate implementation, various techniques have been developed to design controls to deal with specific types of errors that occur frequently in applications such as suppressing leakage to higher excited states~\cite{Economou,MotzoiDRAG}, minimizing crosstalk between qubits~\cite{Quiroz}, and mitigating the effects of resonance offsets~\cite{Glaser2021}. The approach to solving these problems mostly involves finding smooth pulses for two-state systems~\cite{Sugny_STA} or more general multi-qubit systems~\cite{Kestner,Mahesh}. Some of these methods can also be used to replace composite, multi-pulse sequences with a single pulse. Bespoke methods have been developed for specific applications such as generating robust entanglement in trapped ion systems~\cite{Hensinger,Ozeri,Li2022}, robust control of solid-state spin systems~\cite{Zhang2022} and fluxonium qubits~\cite{schuster}, and robust Rydberg gates for atom-based quantum computing~\cite{koch_rydberg}. The last of these methods draws inspiration from past work in robust NMR control~\cite{pulses_broadband}, showing the extensibility of robust quantum control techniques across platforms. 

As discussed, alternative approaches to robust control synthesis involve \emph{shortcuts to adiabaticity} and \emph{geometric control}. For example, the former has been used to ensure the robustness of two-level system controllers to higher order~\cite{Sugny_STA}. On the geometric control front, ~\cite{Barnes} derives single-qubit controls that are robust to noise and general pulse errors. Other research~\cite{Yung2021,Xue_2023} focuses on robust geometric quantum control for gate design, while~\cite{Guerin2020} discusses inverse geometric optimization for robust gate design. Finally,~\cite{Tarn} utilizes geometric control to suppress decoherence under strict control-invariance conditions; more work is needed to soften the condition to \emph{almost} control-invariance~\cite{Willems}. Most of this work has focused on single or two-qubit systems. Geometric control can be extended to $N$-level systems~\cite{Schirmer2004} and quantum computing more generally~\cite{Giunashvili2007}. However, it is not necessarily clear that the robustness properties for two- or three-level systems will translate to higher dimensions.

Most work on geometric quantum control is focused on Hamiltonian engineering for closed systems, in which the underlying manifold on which the dynamics are taking place is the Lie group $U(n)$ or $SU(n)$. Some work considers robust control design for quantum dynamical semigroups~\cite{Alicki}, in particular Lie semigroups, which have a rich algebraic and geometric structure (see, e.g.,~\cite{Joachim}). The exploration of these structures gives rise to useful notions of control for open quantum systems~\cite{Altafini,TicozziEtAl}, and the use of (approximate) symmetries in optimal control of the Lindblad system has yielded some numerical results on the escape chimney of two-level open quantum systems~\cite{CBCR}. Attempts have also been made to extend results from closed quantum systems without drift (which have compact state spaces) to open quantum systems with decoherence, which typically have non-compact state spaces.  When the state space is modeled by a finite-dimensional smooth manifold, and the directions the researcher has direct control over generate (as a Lie algebra) the entire tangent space at a point, the same techniques for controllability are applicable, though it may take an arbitrarily long amount of time to reach a particular state.

On the more theoretical side, recent work has revisited the single spin-$\tfrac{1}{2}$ system to investigate deeper questions in quantum control. The work in ~\cite{garon2013time} uses time-optimal control and the Pontryagin Maximum Principle to determine the fastest transfers for a single-qubit system; later work expanded upon this~\cite{Sugny_2level}, and in~\cite{garon2013time}, researchers found that the global phase of the system modifies the optimal control path. That is, while the global phase is not directly measurable, the theoretical optimization landscape can be altered by modifying this degree of freedom. This happens because optimization criteria such as Eq.~\eqref{eq:Pontryagin} involve quantities that do not remain invariant under the global phase. The work in ~\cite{bian2019experimental} experimentally verifies these results. Along similar lines, the global phase can be manipulated to alleviate the classical limitation between fidelity and its robustness~\cite{CDC_phase}. However, most of these results are limited to closed, single-qubit systems with states modeled by wavefunctions.  Further investigation of these systems and extensions to more complex systems is necessary. 

There is also growing interest in linking thermodynamics and network science to quantum control.  For example, research presented in ~\cite{bogdan2017multi} shows that quantum spintronic networks exhibit multi-fractal behavior, which is linked to the thermodynamic potential of free energy.  While some efforts have been made to develop quantum thermodynamics~\cite{beretta1984quantum,dann2020quantum,geusic1967quantum,halpern2018quantum,halpern2019quantum,kosloff2013quantum,lloyd1988black,scully1999quantum}, more research needs to be done to develop this blossoming field and connect it to multi-fractality.  Control of multi-fractal quantum networks itself is also an open area of research.

Finally, in terms of the application space, many applications would benefit from progress in robust quantum control from cold atoms, quantum gas microscopes~\cite{bakr,sherson} and atoms trapped in optical tweezers~\cite{3Darrays,manybodyrev}, to the superconducting qubit architectures being developed by IBM, Google, Microsoft, and many others.  Private quantum-control-specific companies are also joining these efforts by contributing machine learning techniques to characterize and combat noise sources~\cite{Ball2020,biercuk,Kelly2014}.  Further integration of these promising technologies into the commercial and research spheres requires the development of robust control protocols.  As such, there is a growing need to bridge the gap between quantum physics and control theory.  

\section{Conclusion}\label{sec:conclusion}

Robust control of quantum systems involves the extension of tools from classical robust control and geometric control, as well as employing uniquely quantum ideas, such as the study of the Lindbladian of an open quantum system. This survey is meant to serve as a bridge between classical control theorists and quantum physicists, with an emphasis on what has been done most recently in the field. There are other areas not discussed in depth in this survey, such as quantum error correction~\cite{Benedetti2019,Flurin2018,Fowler2012,Knill2004,Terhal2015} and machine learning \cite{Mavadia2017,Torlai2018}, which are also employed in the study of robust quantum systems. There is a vast quantity of open problems in robust quantum control, and some major areas of research include determining the applicability of classical robust control methods in the quantum regime and developing new ideas of how to quantify quantum robustness. From extending notions of robustness from linear classical control (Sec.~\ref{sec:basic}) to Hamiltonian and Lindbladian systems, to the practical application of robust quantum control (Sec.~\ref{sec:struct}), solving problems in robust quantum control requires an interdisciplinary and collaborative effort. We hope this survey facilitates this effort and inspires future research on robust quantum control.

\begin{ack}                            
This work was partially supported by NSF IRES Grant \#1829078. 
\end{ack}

\bibliographystyle{plain}
\bibliography{bib/bibliography}

\end{document}